\documentclass[aip,pre,twocolumn,floatfix,superscriptaddress,10pt]{revtex4}
\usepackage{graphicx}
\usepackage{amsmath}
\usepackage{color}
\begin{document}

\title{Co-non-solvency: Mean-field polymer theory does not describe polymer collapse transition in a mixture of two competing good solvents}

\author{Debashish Mukherji}
\affiliation{Max-Planck Institut f\"ur Polymerforschung, Ackermannweg 10, 55128 Mainz Germany}
\author{Carlos M. Marques}
\affiliation{Max-Planck Institut f\"ur Polymerforschung, Ackermannweg 10, 55128 Mainz Germany}
\affiliation{Institut Charles Sadron, Universit\'e de Strasbourg, CNRS, Strasbourg, France}
\author{Torsten Stuehn}
\affiliation{Max-Planck Institut f\"ur Polymerforschung, Ackermannweg 10, 55128 Mainz Germany}
\author{Kurt Kremer}
\affiliation{Max-Planck Institut f\"ur Polymerforschung, Ackermannweg 10, 55128 Mainz Germany}



\begin{abstract}
Smart polymers are a modern class of polymeric materials that often exhibit unpredictable behavior in mixtures of solvents. 
One such phenomenon is co-non-solvency. Co-non-solvency occurs when two (perfectly) miscible and competing good solvents, for a given polymer, 
are mixed together. As a result, the same polymer collapses into a compact globule within intermediate mixing ratios. 
More interestingly, polymer collapses when the solvent quality remains good and even gets increasingly better by the 
addition of the better cosolvent. This is a puzzling phenomenon that is driven by strong local concentration fluctuations. 
Because of the discrete particle based nature of the interactions, Flory-Huggins type mean field arguments become unsuitable. 
In this work, we extend the analysis of the co-non-solvency effect presented earlier 
[Nature Communications 5, 4882 (2014)]. We explain why co-non-solvency is a generic phenomenon that can be understood 
by the thermodynamic treatment of the competitive displacement of (co)solvent components. 
This competition can result in a polymer collapse upon improvement of the solvent quality.
Specific chemical details are not required to understand these complex conformational transitions. 
Therefore, a broad range of polymers are expected to exhibit similar reentrant coil-globule-coil transitions in 
competing good solvents.
\end{abstract}

\maketitle

\section{Introduction}
\label{intro}

The microscopic understanding of smart polymer conformations in a mixture of solvents is scientifically challenging \cite{mukherji14natcom} 
and, at the same time, possesses great technological implications that span over a broad range of disciplines \cite{cohen10natmat,ward11poly,sissi14natcom,sissi14mac}. 
Therefore, establishing the links between smart polymer conformations and its specific interactions is a key to develop any 
fundamental understanding of their solubility. Examples of the most commonly known smart polymers include: 
poly(N-isopropylacrylamide) (PNIPAm), poly(N-isopropylmethacrylamide) (PNIPMAm), poly(N,N-diethylacrylamide) (PDEAm), 
poly(N-vinlycaprolactam) (PVCL), and poly(acryloyl-L-proline methyl ester) (PAPOMe). 
When some of these polymers are dissolved in a mixture of solvents, such as aqueous alcohol solutions, they 
show a puzzling coil-globule-coil scenario \cite{schild91mac,zhang01prl,walter12jpcb,koj12jpsb,tanaka08prl,mukherji13mac}. 
This interesting phenomenon is termed as co-non-solvency. 

Theoretical understanding of these complex phenomena is mostly restricted to a limited number of computer simulation
studies \cite{mukherji14natcom,walter12jpcb,walter10fluid,tucker12mac,hayda13mac,mukherji13mac}, which usually deal with
chemically specific details \cite{walter12jpcb,walter10fluid,tucker12mac,mukherji13mac}. 
Moreover, these simulations require careful parameterization of force fields that can be cumbersome, if their are rather delicate differences 
in interactions. However, in this context, if a physical phenomenon can be characterized within a universal concept, 
such that the chemical details only contribute to a pre-factor, then the correct physics can be captured with a rather 
simple generic model. The use of generic schemes has several advantages; 1) the parameter space is 
not restricted to a specific system, unlike the all-atom simulations, and a broad range of systems can be 
represented within a unified simulation protocol, 2) the time scale of simulations are not of concern, 
and 3) because of the absence of any competing energy scales, one does not need an advanced molecular 
dynamics scheme. In this context, we have recently shown that the complexity of smart polymers can be 
captured within a generic model \cite{mukherji14natcom}. Using a simple model we could 
quantitatively capture the reentrant coil-globule-coil scenario of PNIPAm \cite{schild91mac,zhang01prl,walter12jpcb,koj12jpsb} 
and PAPOMe \cite{hiroki01polymer} in aqueous methanol mixtures. Our analysis suggested that when two competing and 
individually good solvents are mixed together, because of the preferential binding of the better of the two (co)solvents
with the polymer, it collapses within the intermediate solution compositions. 
At a low cosolvent concentration, the cosolvent molecules can bind to two distinctly far monomers 
forming bridges and leading to polymer collapse. When the concentration of the better
cosolvent is increased, they decorate the whole polymer and the polymer opens up.
These results are in good agreement with the simulations incorporating all atom details \cite{mukherji13mac}
and experiments \cite{walter12jpcb}. This coil-globule-coil scenario is a generic effect and 
many polymers are expected to exhibit similar behavior as long as one of the solvents is significantly better than the other. 
Thus the behavior is not strictly restricted to the so called smart polymers exhibiting a LCST. In Table \ref{tab:cns} 
we present a list of polymers that show co-non-solvency effect. It is interesting to note that well known standard 
polymers, such as poly(ethylene oxide) (PEO) and polystyrene, also show co-no-solvency \cite{lund04mac,wolf78macchem}.
\begin{table*}[ht]
\caption{A table listing various polymer systems that show co-non-solvency effect when solvated in 
their respective mixture of solvents.}
\begin{center}
\begin{tabular}{|c|c|c|}
\hline
Polymer (${p}$) & Solvent (${s}$) & Cosolvent (${c}$) \\\hline
\hline
Poly(N-isopropylacrylamide) (PNIPAm) \cite{schild91mac,zhang01prl,walter12jpcb,koj12jpsb,winnik90mac} & Water &  Methanol, Tetrahydrofuran, \\
& & or 1,4-dioxane \\
Poly(acryloyl-L-proline methyl ester) (PAPOMe) \cite{hiroki01polymer} & Water & Methanol \\
Poly(ethylene oxide) (PEO) \cite{lund04mac} & Water & N,N-dimethylformamide \\
Polystyrene \cite{wolf78macchem} & N,N-dimethylformamide & Cyclohexane \\
Poly(vinyl alcohol) \cite{ohkura92pol} & Water & Dimethyl sulfoxide \\
Poly(2-(methacryloyloxy)ethylphosphorylcholine) & & \\ 
(PMPC) \cite{kiritoshi02,kiritoshi03} & Water & Methanol, Ethanol, or Iso-propanol \\
\hline
\end{tabular}  \label{tab:cns}
\end{center}
\end{table*}
Another example may include polymeric semiconductors \cite{lemmer2013} that show 
anomalous viscosity with solvent composition, suggesting a change in polymer conformation.  

One of the most intriguing aspects of the co-non-solvency effect is that the solvent quality becomes 
increasingly better by the addition of the better cosolvent. Thus the polymer collapses in a good solvent, making the 
solvent quality decoupled from the polymer conformation. This is very striking and against the 
conventional view on polymer solutions. While the atomistic simulations \cite{mukherji13mac} clearly 
demonstrate that the solvent becomes increasingly better by the addition of better cosolvent,
it is still difficult to identify the preferred local coordination and especially the bridging.
In contrast, generic simulations give a clear microscopic 
understanding of this complex phenomenon within a simple simulation protocol \cite{mukherji14natcom}. 

Complementary to that computer simulations give a good microscopic picture of the polymer collapse
transitions, it is also advantageous to devise a general analytical 
theory consistent with the findings known from computer simulations and/or experiments. 
Moreover, because of the complexity of the system interactions, this discrete particle based 
phenomenon can not be explained using a Flory-Huggins type mean-field picture. 
Instead, these complex conformational transitions can be explained within a Langmuir-like 
thermodynamic treatment of competitive displacement of different solvent components onto the polymer \cite{mukherji14natcom}. 

In this work, we revisit the co-non-solvency effect of smart polymers in the mixtures of solvents. 
We extend the analysis of our previous work \cite{mukherji14natcom} to better understand the 
microscopic picture of co-non-solvency. We will present an in-depth argument to show 
that the mean-field theory is highly unsuitable for these systems and the conceptual need 
of a discrete particle-based theory. We also propose a phase diagram to identify the 
conformational states of smart polymers in various bulk solutions and with the change of 
cosolvent concentrations.

The remainder of the paper is organized as follows: in Section \ref{sec:method} the generic 
molecular dynamics simulation details are presented. Results and the theoretical arguments 
are presented in Section \ref{sec:results} and we finally present our conclusions in Section \ref{sec:con}.

\section{Model and Methods}
\label{sec:method}

We start by briefly describing the details of the generic molecular simulations. 
A similar model has been used in our earlier study. A detailed description of model and method 
is presented in Ref.~[\onlinecite{mukherji14natcom}].
Here a polymer $p$ is modeled using the well known bead-spring polymer model \cite{kremer90jcp}. In this model, individual monomers 
of a polymer interact with each other via a repulsive 6-12 Lennard-Jones (LJ) potential (WCA potential). 
Additionally, adjacent monomers in a polymer are connected via a finitely extensible nonlinear elastic potential (FENE).
Here $p-p$ interaction energy is chosen as $\varepsilon_{p} = 1.0\varepsilon$ and the size of the monomer is $\sigma_{p} = 1.0\sigma$. 
All units are expressed in terms of the LJ energy $\varepsilon$, the LJ radius $\sigma$, and the mass $m$ of individual particles. 
This leads to a time unit of $\tau = \sigma \sqrt{m/\varepsilon}$.
The parameters of the potential are such that a reasonably large time step can be chosen, while bond crossing
remains essentially forbidden.

A bead-spring polymer is solvated in mixed solutions composed of two components also modeled as LJ beads,
solvent $s$ and cosolvent $c$, respectively. 
Since the solvent molecules typically are much smaller than the monomers of PNIPAm and/or PAPOMe in aqueous methanol,
we choose the sizes of (co)solvents to be $\sigma_{s/c} = 0.5\sigma$. Note that because of the reduced size of 
(co)solvents, the corresponding number density within the simulation domain should also be adjusted such that the 
overall pressure remains $\sim 40 \varepsilon/\sigma^3$. 
$p-s$ and $p-c$ interactions are chosen such that $c$ is always a better solvent than $s$. In our earlier study \cite{mukherji14natcom},
the default system consisted of a repulsive $p-s$ interaction, while $p-c$ interaction 
was attractive. In this work, we generalize this and investigate the effect of assymetry in interaction energies, when 
both $p-s$ and $p-c$ are attractive with interactions $\varepsilon_{ps}$ and $\varepsilon_{pc}$, respectively. 
Additionally, we impose conditions; (1) $\varepsilon_{ps} < \varepsilon_{pc}$, (2)  $0.5 < \varepsilon_{ps} < 1.0$ and 
$0.5 < \varepsilon_{pc} < 2.5$. Temperature is set to $T=0.5\varepsilon/\kappa_{\rm B}$, where $\kappa_{\rm B}$ is the Boltzmann constant.
This leads to a relative energy scale $\varepsilon_{pc} - \varepsilon_{ps} \le 3\kappa_{\rm B}T$. 
These values are typically comparable to the interaction energy scale for PNIPAm in aqueous methanol. 
Solvent particles always repel each other with a repulsive LJ potential, with $\epsilon_{ij} = 1.0\epsilon$. 
This is a good approximation given that the $p-c$ and $p-s$ interactions are dominant, which will be discussed 
at a later stage. 

The cosolvent mole fraction $x_c$ is varied from 0 (pure $s$ component) to 1 (pure $c$ component).
We consider three different polymer chain lengths $N_l = 10$, $30$ and $100$, solvated in $2.5 \times 10^4$
solvent molecules for $N_l = 10$ \& $N_l = 30$ and $10 \times 10^4$ solvent molecules for $N_l = 100$, respectively.
The equations of motion are integrated using a velocity Verlet algorithm with a time step $\delta t = 0.005\tau$
and a damping coefficient $\Gamma = 1.0\tau^{-1}$ for the Langevin thermostat.
The initial configurations are equilibrated for typically several $10^5 \tau$, depending on the chain length, which is at least an order of magnitude
larger than the relaxation time in the system. After this initial equilibration averages are taken over another $10^4 \tau$
to obtain observables, especially gyration radii $R_{\rm g}$, chemical potentials $\mu_{p}$ of the polymer and the bridging fractions
of cosolvents $\phi_{\rm B}$.

\section{Results and discussions}
\label{sec:results}

\subsection{Co-non-solvency: A brief overview}

In our previous paper \cite{mukherji14natcom}, we provided a possible explanation for the experimentally observed 
co-non-solvency effect of smart polymers in aqueous mixtures \cite{schild91mac,zhang01prl,walter12jpcb,koj12jpsb}. 
\begin{figure}[ptb]
\includegraphics[width=0.46\textwidth,angle=0]{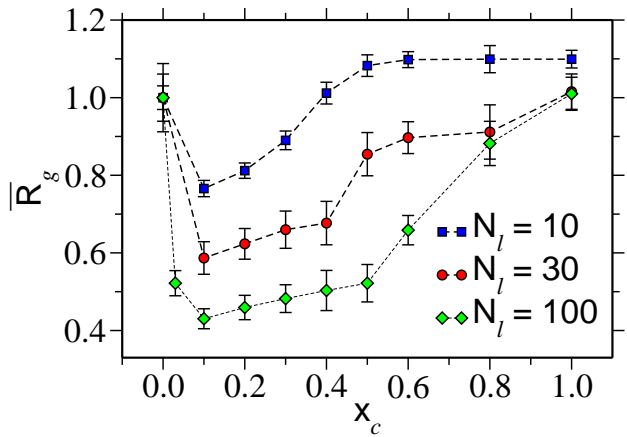}
\caption{(color online) Normalized radius of gyration ${\bar R}_{\rm g} = R_{\rm g}/R_{\rm g} (x_c = 0)$ 
as a function of cosolvent molar concentration $x_c$ for three different chain lengths $N_l$.
Results are shown for the default system taken from Ref.~[1]. The error bars are the standard deviations calculated from six independent simulations.
The lines are drawn to guide the eye.
\label{fig:rg_N}}
\end{figure}
In Fig.~\ref{fig:rg_N} we show the normalized radius of gyration ${\bar R}_{\rm g} = R_{\rm g}/R_{\rm g} (x_c = 0)$
as a function of cosolvent molar fraction $x_c$. The data is shown for three different $N_l$.
It can be appreciated that just by adding a small fraction of the better of the two solvents, the polymer collapses into 
a compact globule structure. As discussed in the introduction, this reentrant collapse and swelling transition is 
facilitated by the preferential binding of cosolvent components with the polymer. The initial collapse is due to 
the formation of bridges that the cosolvent molecules form by binding two monomers that can be 
distinctly far along the backbone of a polymer, while the reopening at higher concentrations is
due to the increased decoration of the polymer by cosolvent molecules. Therefore, we can identify two kinds of cosolvents:
fraction $\phi_{\rm B}$ of bridging cosolvents that bind to two monomers and a fraction $\phi$ of cosolvents 
that are only bound to one monomer. Note that other than $\phi_{\rm B}$ and $\phi$, there are 
a large fraction of free cosolvents that are present in the bulk solution and usually are required to 
maintain solvent equilibrium.

It is also interesting to observe an inverse system size effect in the reopening transition, as observed in Fig. \ref{fig:rg_N}.
While the initial collapse ($x_c < 0.1$) is reminiscent of a first-order-like collapse, the reopening (for $x_c > 0.5$) 
is rather smooth even for longer chain lengths. This is contrary to the knowledge of critical phenomena. Thus 
indicating that this transition is not a phase transition in a true thermodynamic sense. This aspect will be discussed 
at a later stage of this manuscript. Furthermore, the cosolvent driven first-order-like collapse for $x_c \le 0.1$ 
is reminiscent of the temperature induced first order transition in PEO \cite{jeppesen95epl}.

\begin{figure}[ptb]
\includegraphics[width=0.4\textwidth,angle=0]{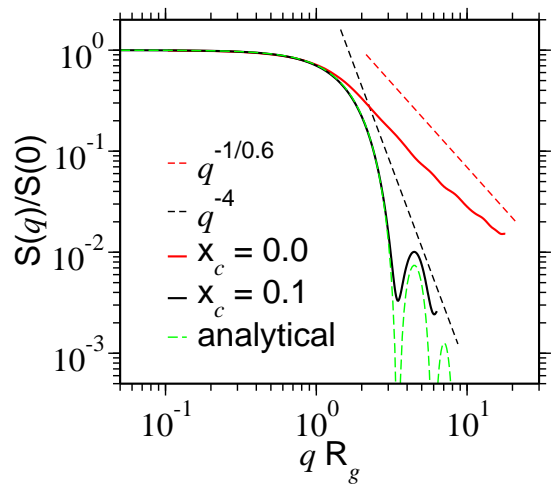}
\caption{(color online) Static structure factor $S(q)$ for a chain length $N_l = 100$ and for two different 
mole fractions $x_c$. A power law of $q^{-1/\nu}$ with $\nu = 0.6$ shows an extended (good solvent) conformation and 
$q^{-4}$ supports a compact globule structure. For comparison, we have also plotted the analytical scattering function of a
sphere.
\label{fig:sofk}}
\end{figure}

To further quantify the collapse facilitated by bridging cosolvents, we calculate the static structure factor $S(q)$ \cite{jeppesen95epl,higginsbook},
\begin{equation}
S(q) = {\frac {1}{N_l}} \left< \left|\sum_{i} e^{\left[i {\bf q}\cdot {\bf R}_i \right]} \right|^2\right>.
\label{eq:sofq}
\end{equation}
In Fig.~\ref{fig:sofk} we present $S(q)$ for $N_l = 100$.
As expected, a power law  well approximated by $q^{-1/0.6}$ is observed for $x_c = 0$ (pure solvent), a signature characteristics of an extended
coil structure. For $x_c = 0.1$, the polymer collapses into a compact globule, with $\sim 60\%$ decrease in $R_{\rm g}$ with 
respect to its original extended $R_{\rm g}$ at $x_c = 0$ (see Fig.~\ref{fig:rg_N}), as shown by a
prominent scaling law $q^{-4}$ in Fig.~\ref{fig:sofk} \cite{higginsbook}. Note that $R_{\rm g}$ for $x_c =0.1$ is 
slightly larger than the equivalent $R_{\rm g}$ when a polymer collapses because of pure depletion effects. 
This is due to the fact that a collapsed polymer also contains interstitial bridging cosolvent and their sizes
contribute towards a slightly larger compact globule.

Preferentiability is required for the observation of the co-non-solvency effect. Therefore, conformations of polymers
in mixed solvents are intimately linked to the asymmetry in $p-c$ and $p-s$ interactions 
$\varepsilon_{pc} - \varepsilon_{ps}$. 
\begin{figure*}[ptb]
\includegraphics[width=0.79\textwidth,angle=0]{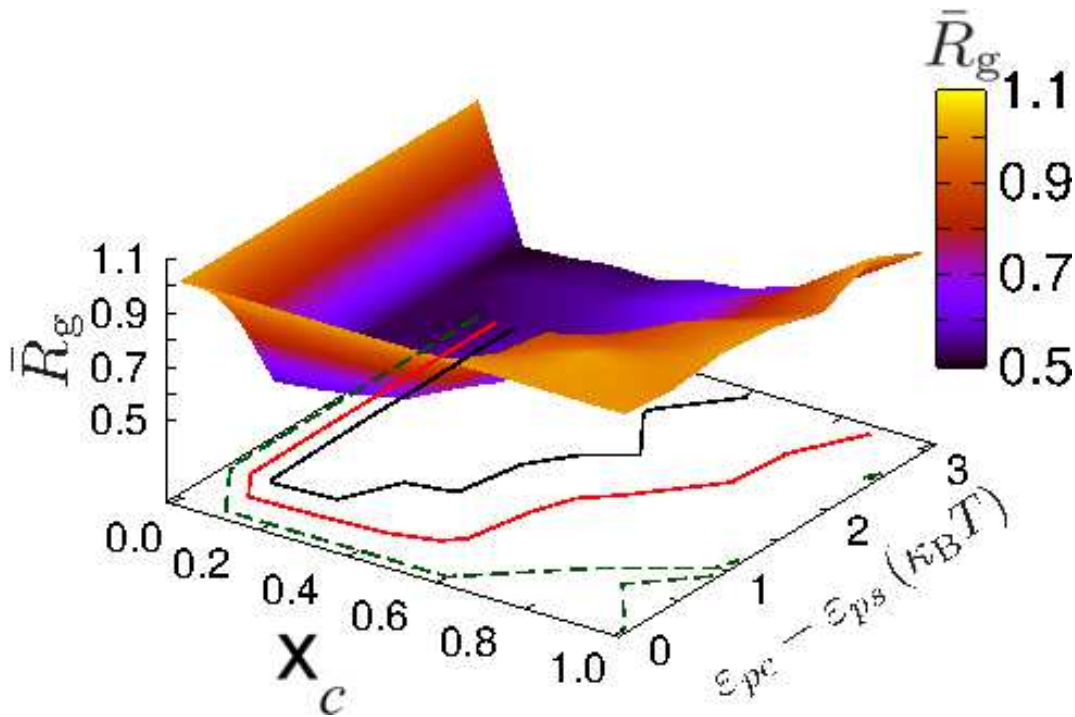}
\caption{(color online) A sketch of the phase diagram showing the change in normalized radius of gyration
${\bar R}_{\rm g} = R_{\rm g}/R_{\rm g} (x_c = 0)$ with the varying cosolvent molar concentration $x_c$ and
the relative interaction strengths $\varepsilon_{pc} - \varepsilon_{ps}$. Results are shown for chain length $N_l = 30$.
In the contour plots, the area bound by the black curve represents maximum collapse with ${\bar R}_{\rm g} < 0.6$.
The red contour curve represents the boundary when polymer goes from globule-coil (or vice-versa) by either changing 
$x_c$ at a constant $\varepsilon_{pc} - \varepsilon_{ps}$ or by changing $\varepsilon_{pc} - \varepsilon_{ps}$ at constant $x_c$. 
The region outside the dashed green curve shows maximum extension of the chain with ${\bar R}_{\rm g} \ge 1.0$. 
Prominent kink in the black contour curve is due to the error bar associated with that data points.
\label{fig:phase_d}}
\end{figure*}
In Fig.~\ref{fig:phase_d} we present a unified picture of polymer 
conformation with changing $\varepsilon_{pc} - \varepsilon_{ps}$ at different $x_c$.
It can be appreciated that, for $\varepsilon_{ps} = \varepsilon_{pc}$, co-non-solvency is not observed and 
the polymer remains in a coil conformation. Only when $\varepsilon_{pc} - \varepsilon_{ps} > 0.25 \kappa_{\rm B} T$, 
does the polymer exhibit a coil-globule-coil-like scenario. More interestingly, the larger the difference $\varepsilon_{pc} - \varepsilon_{ps}$ 
the smoother the re-opening transition at larger $x_c$ values. This is not surprising given that for a stronger 
$\varepsilon_{pc}$, $\phi_{\rm B}$ has stronger binding, thus leading to a more stable semi-collapsed conformation.
Thermodynamically, the energy density in the solvation shell can be increased by increasing the $p-c$ interaction strength. 
Increasing energy density by a factor of two will approximately act in a similar manner as that of a polymer of twice 
$N_l$. Interestingly enough, increased $\varepsilon_{pc}$ and/or $N_l$ has the same effect on the overall polymer conformation.

We also want to point out that, for $\varepsilon_{pc} >> \varepsilon_{ps}$, 
polymer collapse will occur close to $x_c \to 0$. 
A more prominent representation of the region $0.0 < x_c < 0.1$ will require fine grids and 
systematic scanning of the concentrations within the range $0 < x_c < 0.1$. 
Here, however, while the initial collapse is always first order like, 
the re-opening has much stronger dependence on $\varepsilon_{pc}- \varepsilon_{ps}$. 
Therefore, we abstain from presenting any more details for $0 < x_c < 0.1$.

The most interesting aspect of this reentrant transition is that even when the solvent quality becomes better and better by the 
addition of the better good solvent, the polymer collapses in good solvent. This makes the polymer conformation decoupled from 
the solvent quality and only dictated by the preferential coordination of cosolvent with polymer. This particle based 
phenomenon can not be explained within a mean-field type approach.

Before describing an analytical theory, we briefly want to comment on the suitability of the generic simulation 
protocol to study the complexity of smart polymers. One important aspect of smart polymers, such as PNIPAm, PVCL, and 
PAPOMe, is their thermal responsiveness. These polymers remain in a coil configuration at low temperatures, while 
collapsing into a compact globule at high temperatures, thus presenting a lower critical solution temperature (LCST).
In this context, it is worth mentioning that the generic schemes do not present LCST and co-non-solvency 
can be studied at one fixed temperature over full range of $x_c$. Moreover, the co-non-solvency effect is not 
necessarily restricted to the ``so called" smart polymers exhibiting LCST. Therefore, a broad range of polymers are 
expected to show a similar reentrant scenario, as long as they are dissolved in a mixture of competing good solvents.  
A list of the possible polymer systems that show co-non-solvency is presented in Table~\ref{tab:cns}. Standard polymers, 
such as PEO in aqueous DMF \cite{lund04mac} and polystyrene in a mixture of N,N-dimethylformamide (DMF) and cyclohexane \cite{wolf78macchem},
also show co-non-solvency \cite{lund04mac}. Another example includes polymeric semiconductors. 
It has been observed that the solution viscosity of polymeric semiconductors can display a non-monotonic dependency on cosolvent concentration,
indicating conformational change \cite{lemmer2013}. Therefore, we speculate that many more polymers, such as polycarbonate or 
polypropylene, may also exhibit a similar reentrant transition in appropriate competing good solvents.

\begin{figure}[ptb]
\includegraphics[width=0.37\textwidth,angle=0]{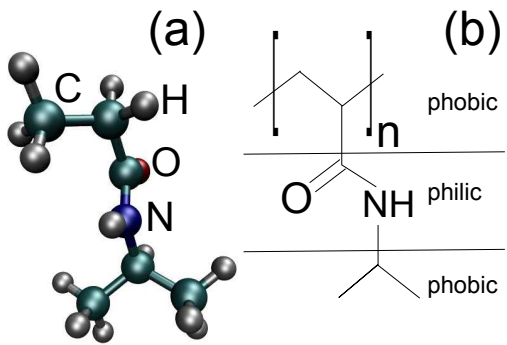}
\caption{(color online) Part (a) shows simulation snapshot representing a monomer of PNIPAm. Hydrogen atoms is rendered in steel, 
green spheres are Carbon atoms, blue sphere is Nitrogen, and the Oxygen is rendered in red. Part (b) represents a 
schematic representation of NIPAm monomer.
\label{fig:cart}}
\end{figure}

Another aspect is that the NIPAm monomer has a hydrophilic part and two hydrophobic parts, as shown in the 
schematic Fig.~\ref{fig:cart}. 
It is generally believed that methanol molecules bind to the hydrophilic part and thus push away water molecules towards the hydrophobic part, 
leading to polymer collapse. In a generic simulation, however, the monomer is represented by a sphere, thus 
eliminating any effects due to hydro(phob/phil)icity within the model. Our simulations \cite{mukherji14natcom,mukherji13mac} suggests that the 
only dominant interaction is the preferential coordination of methanol around the NIPAm monomer (see Fig. 3 of Ref. [\onlinecite{mukherji13mac}]).
The collapse is initiated by bridging and not by any hydrophobic effects that may occur due to solvent interaction with the alkane 
backbone. Chemical details do not play any role in describing this reentrant transition. 
In this context, tuning specific (co)solvent-polymer interactions, a whole new class of heteropolymers can also 
exhibit co-non-solvency and related phenomena \cite{future}.

\subsection{Co-non-solvency: A simple analytical approach}

\subsubsection{Why mean-field theory is inappropriate to describe co-non-solvency}

\begin{figure}[ptb]
\includegraphics[width=0.49\textwidth,angle=0]{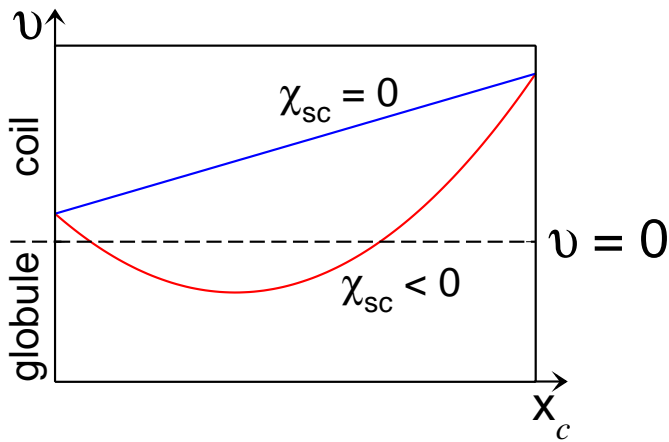}
\caption{(color online) A schematic representation of the polymer excluded volume v as a function of
cosolvent mole fraction $x_c$. The curve shows that the interaction parameter $\chi_{sc}$ between solvent-cosolvent is a key factor to
exhibit a swelling-collapse-swelling scenario. When $\chi_{sc} = 0$ the polymer remains swollen.
\label{fig:fh_exvol}}
\end{figure}

When a polymer with chain length $N_l$ at volume fraction $\phi_p$ is dissolved 
in a mixture of two components $s$ and $c$, respectively, the standard 
Flory-Huggins energy ${\mathcal F}_{\rm FH}$ of polymer solutions reads \cite{degennesbook,desclobook},
\begin{eqnarray}
\frac {{\mathcal F}_{\rm FH}}{\kappa_{\rm B} T} &=& \frac {\phi_{p}}{N_l}\ln{\phi_{p}} + x_c \left(1 - {\phi_{p}} \right) \ln\left[x_c \left(1 - {\phi_{p}} \right)\right] \nonumber\\
&+& \left( 1 - x_c \right) \left(1 - {\phi_{p}} \right) \ln\left[\left( 1 - x_c \right) \left(1 - {\phi_{p}} \right)\right] \nonumber\\
&+& \chi_{ps} {\phi_{p}} \left( 1 -x_c \right) \left(1 - {\phi_{p}} \right) \nonumber\\
&+& \chi_{pc} {\phi_{p}} x_c \left(1 - {\phi_{p}} \right) \nonumber\\
&+& \chi_{sc} x_c \left( 1 -x_c \right) \left(1 - {\phi_{p}} \right)^2.
\label{eq:fh_freng}
\end{eqnarray}
Here, the first three terms represent the entropy of mixing and the last three terms deal with interactions between
different components $i$ and $j$ via $\chi_{ij}$. Expanding Eq.~\ref{eq:fh_freng} to the second order gives a
direct measure of the excluded volume ${\mathcal V}$ of the polymer \cite{degennesbook,desclobook};
\begin{eqnarray}
{\mathcal V} &=& 1 - 2 \left( 1 - x_c \right) \chi_{ps} - 2x_c \chi_{pc} + 2 x_c \left( 1 - x_c \right) \chi_{sc},
\label{eq:fh_exvol}
\end{eqnarray}
where $\chi_{ps}$ and $\chi_{pc}$ are the Flory-Huggins interaction parameters between $p-s$ and $p-c$, respectively.
The factor $\chi_{sc}$ is the parameter of $s-c$ interaction. When both solvent and cosolvent are good solvents,
$\chi_{ps} < 1/2$ and $\chi_{pc} < 1/2$ ~\cite{schild91mac}. Using the first two terms of Eq.~\ref{eq:fh_exvol}, we find a
linear variation of ${\mathcal V}$ with $x_c$ for the cases of non-interacting $s$ and $c$ (i.e. $\chi_{sc}=0$), as shown by the 
blue line in Fig.~\ref{fig:fh_exvol}. It is also clear from Fig.~\ref{fig:fh_exvol} that only when $\chi_{sc} < 0$ can 
${\mathcal V}$ become negative, opening the possibility for the coil-to-globule-to-coil conformation changes typical of co-non-solvency. 
It has been noticed early \cite{schild91mac} that for common solvent mixtures where co-non-solvency effects 
are observed, such as water-alcohol mixtures,   $\chi_{sc} > 0$, thus precluding any explanation based on a mean-field, Flory-Huggins type of analysis.

\begin{figure}[ptb]
\includegraphics[width=0.43\textwidth,angle=0]{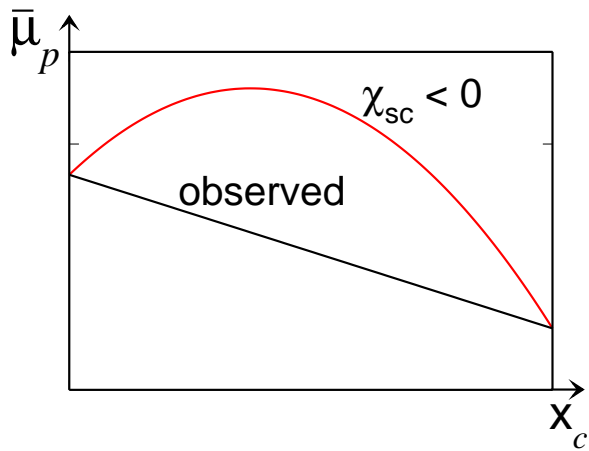}
\caption{(color online) A schematic representation of the shift in chemical potential ${\overline{\mu}}_p$ as a function of cosolvent mole fraction $x_c$. 
\label{fig:fh_mu}}
\end{figure}

Furthermore, within the mean-field picture described in Eq.~\ref{eq:fh_freng}, 
one can get the expression for the shift in chemical potential of polymer $\bar{\mu}_{p}$ for $\phi_p \to 0$,
\begin{eqnarray}
\bar{\mu}_{p} \left(\phi_p \to 0 \right) &=& \frac {\partial {{\mathcal F}_{\rm FH}}}{\partial \phi_{p}}{\bigg{|}_{\phi_p \to 0}} \nonumber\\
&=& const - x_c \ln x_c - \left(1 - x_c\right) \ln {\left(1 - x_c\right)} \nonumber\\
&+& {\left(1 - x_c\right)} \chi_{ps} + x_c \chi_{pc} \nonumber\\ 
&-& 2 x_c \left( 1 - x_c \right) \chi_{sc}.
\label{eq:fh_chempot}
\end{eqnarray}
In Fig.~\ref{fig:fh_mu}, we present a schematic representation of $\bar{\mu}_{p}$ as expected from Flory-Huggins picture 
described in Eq.~\ref{eq:fh_chempot}. 
Here $\bar{\mu}_{p} (x_c = 1) < \bar{\mu}_{p} (x_c = 0)$ because alcohol is a better solvent compared to water.
Consistently with the behavior of ${\mathcal V}$ presented in Fig.~\ref{fig:fh_exvol}, 
$\bar{\mu}_{p}$ for $\chi_{sc} < 0$ displays a hump for intermediate mixing ratios where the solvent quality 
goes from good to poor to good again (see red curve in Fig.~\ref{fig:fh_mu}). 
However, in our simulations \cite{mukherji14natcom,mukherji13mac} not only is $\chi_{sc} = 0$, but we also measure 
a chemical potential trend similar to the black schematic curve in Fig.~\ref{fig:fh_mu}. Thus 
the solvent quality remains good in the whole composition range and, in-fact, it even becomes increasingly better by the addition of 
the cosolvent. By the analysis of Eq.~\ref{eq:fh_chempot}, it can be seen that a similar trend as the black curve of Fig.~\ref{fig:fh_mu} can be
obtained from the mean-field picture when $\chi_{sc} >> 0$. However, this can only be obtained at the nonrealistic cost of driving the system towards 
solvent phase separation. This further confirms the incapability of mean-field theory to capture the reentrant 
co-non-solvency effect in polymeric systems.

The mean-field picture also suggests that the strength of $s-c$ interaction should be dominant over
$p-s$ and $p-c$ interactions to observe this reentrant transition. Moreover, if the mean-field theory is 
sufficient to understand this reentrant coil-globule-coil transition then the analysis of the bulk solution property, that can 
easily be calculated using molecular simulations, should also show a preferred $s-c$ coordination 
over $p-s$ or $p-c$ coordination. 

\begin{figure}[ptb]
\includegraphics[width=0.46\textwidth,angle=0]{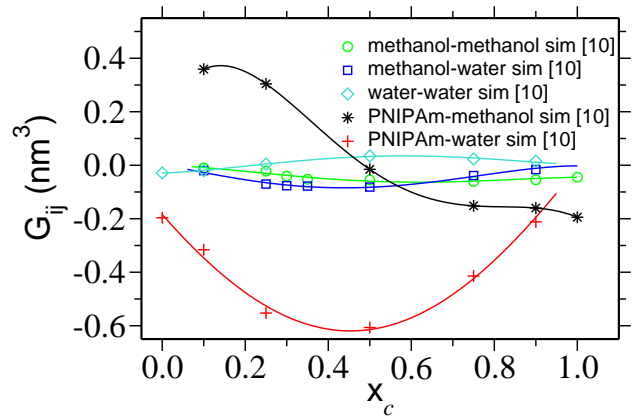}
\caption{(color online) Kirkwood-Buff integral $G_{ij}$ between different solution components as a function of cosolvent
molar fraction $x_c$. Lines are the polynomial fits to the data that are drawn to guide the eye. 
The data was obtained from the semi-grand canonical simulations incorporating all-atom 
details \cite{mukherji13mac}. For pure solvent at $x_c = 0.0$ and pure cosolvent at $x_c = 1.0$, individual 
coordinations $G_{pc}$ and $G_{ps}$ are undefined, respectively.
\label{fig:kbi}}
\end{figure}

A quantity that best describes the relative intermolecular affinity and/or the interaction strength is the 
fluctuation theory of Kirkwood and Buff (KB) \cite{kb51jcp}. KB theory connects the pair distribution function 
to thermodynamic properties of solutions using the ``so called" KB integrals;
\begin{equation}
{\rm G}_{ij} = 4\pi \int_0^\infty \left[ {\rm g}_{ij}(r) - 1\right] r^2 dr,
\label{eq:kbi}
\end{equation}
where ${\rm g}_{ij}(r)$ is the pair distribution function. 
In Fig.~\ref{fig:kbi} we summarize $G_{ij}$ between different solvent components.
It can be appreciated that, for $0.1 < x_c < 0.5$, $p-c$ coordination is at-least an order of magnitude larger than the 
$G_{ij}$ values between the solvent components in the bulk solution, suggesting that the fraction of cosolvent molecules in close contact with the chain is 
always much larger than its natural, mean-field proportions in the bulk solution. This is contrary to what is know from the analysis based on the 
mean-field theory presented above. Furthermore, the shift in chemical potential $\bar{\mu}$ can be estimated from the 
KB theory. If a polymer $p$ at dilute concentration is solvated in a mixture of solvent $s$ and cosolvent $c$,
$\mu_p$ can be calculated using \cite{rosgen05bio},
\begin{equation}
\left(\frac {\partial {\overline {\mu}}_{p}}{\partial {\rho}_c}\right)_{p,T} = \frac {{\rm G}_{ps} -{\rm G}_{pc}}
{1- {\rho}_c {\left({\rm G}_{cs} -{\rm G}_{cc}\right)}},
\label{eq:chempot}
\end{equation}
where ${\overline {\mu}}_{p} = {\mu}_{p}/\kappa_{\rm B}T$, and $\rho_c$ is the cosolvent number density.
\begin{figure}[ptb]
\includegraphics[width=0.25\textwidth,angle=0]{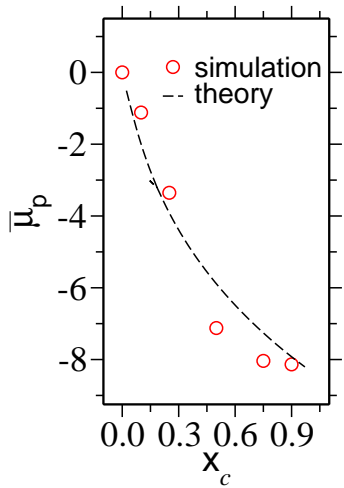}
\caption{(color online) Shift in chemical potential of a single monomer ${\overline {\mu}}_p$ as a
function of methanol mole fraction $x_c$. ${\overline {\mu}}_p$ is calculated by integrating the Eq.~\ref{eq:chempot}. 
$G_{ij}$ are taken from Fig.~\ref{fig:kbi}. Dashed line is plotted according to the Eq.~\ref{eq:chem_pot_an}.
\label{fig:kbi_chem}}
\end{figure}
The change in $\overline{\mu}_{p}$ is shown in Fig.~\ref{fig:kbi_chem}. Data clearly show the trend known for the case when
$\chi_{sc} = 0$, suggesting that the solvent quality becomes better and better by addition of the better (co)solvent.
This decoupling between solvent quality and the polymer conformation is contrary to the conventional understanding from 
mean-field predictions. Additionally, this conformational transition of polymer in mixtures of 
competing good solvents is not a phase transition in true thermodynamics sense and is only dictated by the 
preferential adsorption of one of the (co)solvents. Therefore, an analytical description is needed that can 
incorporate the concept of competitive adsorption by taking into account the strong deviations of local concentration 
from mean-field values. We will address this in the following section. 

\subsubsection{Competitive adsorption of the cosolvent as a model for co-non-solvency}

\begin{figure}[ptb]
\includegraphics[width=0.46\textwidth,angle=0]{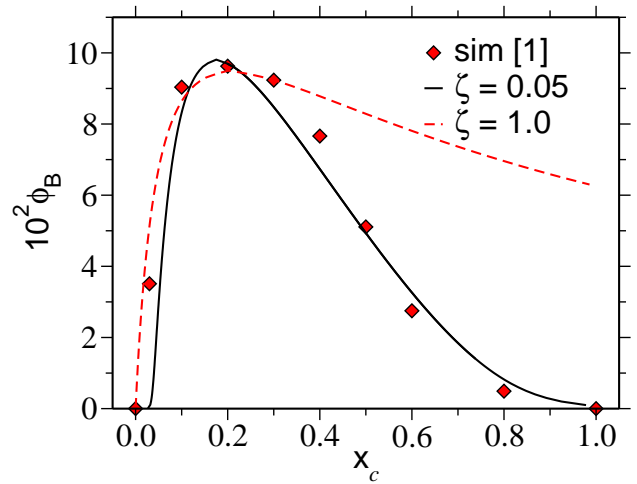}
\caption{(color online) Bridging fraction of cosolvents $\phi_{\rm B}$ as a function of cosolvent mole fraction $x_c$
for $N_l = 100$. The data corresponding to red $\diamond$ is the direct calculation of $\phi_{\rm B}$ from the simulation trajectory.
The prediction of analytical theory from Eq.~\ref{eq:analytic} is plotted for two different $\zeta$ parameters. Here $\zeta = 0.05$
corresponds to the translational entropic term corrected with a loop contribution, as observed earlier \cite{mukherji14natcom}. 
\label{fig:bridge_frac}}
\end{figure}

We have recently proposed that the polymer collapse in co-non-solvency phenomena can simply be understood as 
the result of the attractive interactions induced by cosolvent molecules that form a bridge between 
two monomers \cite{mukherji14natcom}. Conformational collapse, at low $x_c$, is thus induced by the increase of such 
bridges, while polymer swelling at larger cosolvent fractions is due to the progressive replacement 
of these bridges by single site cosolvent molecules that are attached to one monomer only. Thus, 
for high enough $x_c$, these non-bridging cosolvent molecules eventually decorate the whole chain backbone 
to facilitate the reopening. In Fig.~\ref{fig:bridge_frac} we show $\phi_{\rm B}$, the fraction of backbone sites 
participating in bridge formation, as a function of $x_c$. Note that $\phi$ and $\phi_{\rm B}$ are cosolvent molecules that are 
directly in contact with the monomers at a distance $2^{1/6}\sigma_{pc} \sim 0.84\sigma$. It can be appreciated that, within the range
$0.1 < x_c < 0.4$, $\phi_{\rm B}$ obtained from the numerical simulations shows a distinct hump that is consistent with the 
range of $x_c$ when the polymer collapses into a (compact) globule and then gradually begins to expand.
To devise a theoretical formulation, we view the polymer as a substrate with $N$ sites exposed to the bulk solution, 
of which $n^s$ sites are occupied by $s$ (solvent) molecules, $n^c$  sites by non-bridging $c$ (co-solvent) molecules 
and $2 n^c_{B}$ sites by bridging $c$ (co-solvent) molecules, with $N = n^s + n^c + 2 n^c_{B}$. 
The observed sequence of collapse and re-swelling of the polymer corresponds to a fast growth of $n^c_{B}$ as $x_c$ increases, 
followed by a displacement of $n^c_{B}$ by $n^c$ for larger $x_c$ values. 
Such a sequence is typical for competitive displacement in adsorption phenomena \cite{hillbook}. 
Our results from numerical simulations for $n^c_{B}$ and $n^c$, 
or alternatively for the fractions $\phi_{\rm B}=n^c_{B}/N$ and $\phi=n^c/N$, are very well described 
by a competitive adsorption model with the following associated free energy density of adsorption for non-bridges and bridges,
\begin{eqnarray}
\frac {\Psi}{\kappa_{\rm B} T} &=& \phi \ln \left(\phi\right) + \zeta \phi_{\rm B} \ln \left(2\phi_{\rm B}\right) \nonumber\\
&+& \left(1 - \phi - 2 \phi_{\rm B}\right) \ln \left(1 - \phi - 2 \phi_{\rm B}\right) \nonumber\\
&-& \mathcal{E} \phi - \mathcal{E}_{\rm B} \phi_{\rm B} - \frac {\mu}{\kappa_{\rm B} T} \left(\phi + \phi_{\rm B}\right),
\label{eq:f_energy}
\end{eqnarray}
with $\mu = \kappa_{\rm B} T \ln(x_c)$ being the chemical potential of the cosolvent in the bulk solvent mixture and the
adsorption energies $\mathcal{E}$ and $\mathcal{E}_{\rm B}$ measure the excess affinities of individual non-bridging 
and bridging cosolvent molecules to the chain monomers. The first three terms in Eq.~\ref{eq:f_energy} express
entropic contributions of the adsorbed bridges and non-bridges to the energy densities, while the two
following terms measures contact energies between the cosolvents bridges and non-bridges with the 
polymer backbone. The unusual pre-factor $\zeta$ is, as discussed later, a consequence of assuming a logarithmic form for the dependence of 
the energy required to make a bridge on the average density of existing bridges. 
This is the case for instance \cite{mukherji14natcom}, if one assumes that in order to make a new bridge 
at density $\phi_{\rm B}$, the chain needs to make a loop of length $\ell=1/\phi_{\rm B}$, with associated penalty $\sim \log \ell \sim \log(1/\phi_{\rm B})$. 

Minimization of Eq.~\ref{eq:f_energy} with respect to $\phi_{\rm B}$ and $\phi$ leads to the 
implicit equation for the bridge density $\phi_{\rm B}(x_c)$,
\begin{eqnarray}
16 {\phi_{\rm B}}^{\zeta} x_c &=& {x_c^*}
\bigg \{ \left( \frac {x_c^*}{x_c^{**}}\right)^{1/2} \left(1 - 2{\phi_{\rm B}} \right) \nonumber\\
&\pm& \sqrt{\left( \frac {x_c^*}{x_c^{**}}\right) \left(1 - 2{\phi_{\rm B}} \right)^{2} - 16 {\phi_{\rm B}}^{\zeta} } \bigg\}^2.
\label{eq:analytic}
\end{eqnarray}
with $x_c^* = \exp({-\mathcal{E}})$ and $x_c^{**} = \exp({-\mathcal{E}_{\rm B}}+2 \ln 2e -\zeta)$ are the characteristic 
concentrations related to the adsorption energies $\mathcal{E}$ and $\mathcal{E}_{\rm B}$ for non-bridges and 
bridges. Fig.~\ref{fig:bridge_frac} shows that this expression describes very well our experimental results, with $\zeta=0.05$. 

\begin{figure}[ptb]
\includegraphics[width=0.34\textwidth,angle=0]{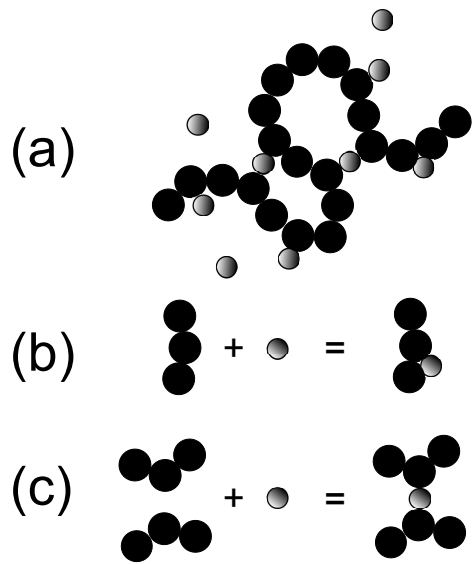}
\caption{(color online) A schematic representation of the chemical reaction described in Eq.~\ref{eq:reac_eq}.
Part (a) describes a typical polymer conformation decorated by non-bridging and bridging cosolvent molecules. 
While part (b) shows that a polymer segment and a cosolvent forms a single adsorbed non-bridging cosolvent, 
part (b) represents two segments and a cosolvent makes a bridge (or a bridging cosolvent).
\label{fig:sch_chem_reac}}
\end{figure}

Eq.~\ref{eq:analytic} can equivalently be derived  by considering the two pseudo chemical 
reactions, 
\begin{eqnarray}
{\rm cosolvent}  +  {\rm empty\ site } \rightleftharpoons {\rm  non-bridge}\nonumber\\
{\rm cosolvent}  +  2 {\rm \ empty\ site } \rightleftharpoons \zeta\ {\rm bridge}.
\label{eq:reac_eq}
\end{eqnarray}
A schematic representation of this reaction is presented in Fig.~\ref{fig:sch_chem_reac}. 
When the solvent and cosolvent interactions with the polymer backbone empty sites are described as pseudo reactions, 
a cosolvent molecule reacts with one empty adsorption site to form one adsorbed non-bridge, 
while it reacts with two empty sites to make $\zeta$ bridges. The associated equilibrium standard mass-action laws can thus be written as
\begin{eqnarray}
\frac{x_c}{x_c^*} &=& \frac{\phi}{1-\phi-2\phi_{\rm B}}\nonumber\\
\frac{x_c}{4 x_c^{**}} &=& \frac{\phi_{\rm B}^\zeta}{(1-\phi-2\phi_{\rm B})^2},
\label{eq:reactions}
\end{eqnarray}
with equilibrium reaction constants $1/x_c^*$  and  $1/x_c^{**}$. 
Note that the reaction equilibrium  concentration $x_c^{**}$ has been, for mathematical convenience, 
defined up to a factor four. Solving the mass-action laws for $\phi_{\rm B}$ gives Eq.~\ref{eq:analytic}.
In this pseudo-chemical language, the factor $\zeta$ describing the effective number of bridges formed 
by the interaction between one cosolvent molecule and the two empty sites of the backbone appears 
as a consequence of assuming a power-law dependence for the equilibrium 
constant of the pseudo-chemical reaction. Note that the actual shape of  Eq.~\ref{eq:analytic}  is quite sensitive to the value of  $\zeta$. 
In particular, the choice $\zeta=1$, corresponding to a standard chemical reaction between free species in solution, 
leads to a prediction that can not describe our data (see red curve in Fig.~\ref{fig:bridge_frac}).

In a previous work \cite{mukherji14natcom}, we argued that a value of $\zeta=0.05$ can be understood by considering loop contributions to the cost of making a bridge. 
When a pure configurational cost for distributing the bridges amongst the possible occupation sites is combined with the entropic 
cost of loop formation, one can write $\zeta=2-m$. Here the critical exponent $m$ can be estimated within a simple scaling argument. 
In this context, one can characterize the loop formation by a partition function of vanishing
end-to-end distance $R_e \rightarrow 0$ \cite{desclobook}, 
\begin{eqnarray}
Z_{N_l} (R_e \rightarrow 0) \propto q^{N_l} {N_l}^{\alpha - 2},
\end{eqnarray}
and the partition function at finite $R_e$ is given by,
\begin{eqnarray}
Z_{N_l} (R_e) \propto q^{N_l} {N_l}^{\gamma - 1}.
\end{eqnarray}
Here $1/q$ is the critical fugacity and the universal exponent $\alpha \cong 0.2$ \cite{desclobook}. From these
two cases one can estimate the free energy barrier to form a loop of length $\ell$ as
$\triangle\mathcal{F}(\ell) = m \kappa_{\rm B} T \ln(\ell)$, with $m = \gamma - \alpha + 1$ being the critical exponent \cite{desclobook}.
Although this gives $m=1.95$ for loop formation in self-avoiding walks, in excellent agreement with our findings, it is worth 
pointing to the fact that our simple analytical description does not address  other possible contributions to bridge formation, 
such as cooperative or other non-trivial entropic effects that might be determinant in the dense chain globule. 

This selective adsorption model provides also for an analytical prediction of the shift in the chemical potential $\mu_p$ as a function of $x_c$,
\begin{eqnarray}
\frac{\mu_p}{\kappa_{\rm B}T} &=& const +  \left(2 - \zeta \right){\phi_{\rm B}} \nonumber\\ 
&-& \ln \bigg\{ 1 + {\phi_{\rm B}}^{1 - \zeta/2} 
\left(\frac {x_c} {x_c^{**}}\right)^{1/2} + \left(\frac {x_c} {x_c^*}\right)\bigg\}.
\label{eq:chem_pot_an}
\end{eqnarray}
Fig.~\ref{fig:kbi_chem} shows a comparison between predictions from Eq.~\ref{eq:chem_pot_an} and the values of the chemical potential obtained from 
Eq.~\ref{eq:chempot}. A very good agreement is obtained by simply inserting into Eq.~\ref{eq:chem_pot_an} the values for $\zeta$ 
and concentrations obtained from the fit of the bridging fraction, further confirming the consistency and validity of our approach.

\section{Conclusion}
\label{sec:con}

We present a comprehensive analysis of the co-non-solvency effect of smart polymers in a mixture of good solvents.
Our results suggest that co-non-solvency is a generic effect that is not restricted to any specific chemical systems. 
Though co-non-solvency has been associated with polymers like PNIPAm, PVL, and/or PAPOMe 
\cite{mukherji14natcom,schild91mac,zhang01prl,walter12jpcb,tanaka08prl,hiroki01polymer}, polymers such as PEO and polystyrene
also exhibit co-non-solvency \cite{lund04mac,wolf78macchem}. Furthermore, this reentrant transition is dictated by the preferential coordination of one of the cosolvents 
with the polymers. More interestingly, even when the chain collapses, the solvent quality becomes increasingly better. This makes the 
solvent quality disconnected from the conformation of the macromolecules. This discrete particle-based phenomenon can 
not be explained within a mean-field theory. Instead it can be explained using a thermodynamic treatment of 
a simple selective adsorption picture and a generic simulation protocol.
Therefore, this work presents a unified theoretical and computational framework, which can pave the way 
for a more generic understanding of polymeric solubility in mixtures of solvents. 

\section{Acknowledgment} We thank Alexander Grosberg, Guojie Zhang and Kostas Daoulas for many stimulating 
discussions and J\"org Stellbrink for bringing Ref.~[\onlinecite{lund04mac}] to our attention. 
C.M.M. acknowledges Max-Planck Institut f\"ur Polymerforschung for hospitality 
where this work was performed. We thank Aoife Fogarty, Robinson Cortes-Huerto, and Debarati Chatterjee 
for critical reading of the manuscript. 
Simulations are performed using ESPResSo++ molecular dynamics package \cite{espresso} and one snapshot in
this manuscript are rendered using VMD \cite{schulten}.

\end{document}